\documentstyle[pre,aps,multicol,epsfig]{revtex}

\begin{document}
\draft

\title{Persistence in the Zero-Temperature Dynamics of the Random
Ising Ferromagnet on a Voronoi-Delaunay lattice}
\author{F.W.S. Lima$^1$ , R. N. Costa Filho$^2$, U. M. S.
Costa$^{2}$ } \address{Departamento de F\'\i sica, Universidade
Estadual Vale do Acara\'u, 62040-370, Sobral, Cear\'a, Brazil}
\address{Departamento de F\'\i sica, Universidade Federal do Cear\'a,
60451-970, Fortaleza, Cear\'a, Brazil}

\date{\today}
\maketitle

\begin{abstract}
The zero-temperature Glauber dynamic is used to investigate the
persistence probability  $P(t)$ in the randomic two-dimensional
ferromagnetic Ising model on a Voronoi-Delaunay  tessellation. We
consider the coupling factor $J$ varying with the distance $r$
between the first neighbors to be $J(r) \propto e^{-\alpha r}$,  with
$\alpha \ge 0$. The persistence probability of spins flip, that does
not depends on time $t$, is found to decay to a non-zero value
$P(\infty)$ depending on the parameter $\alpha$. Nevertheless, the
quantity $p(t)=P(t)-P(\infty)$ decays exponentially to zero over long
times. Furthermore,  the fraction of spins that do not change at a
time $t$ is a monotonically increasing function of the parameter
$\alpha$. Our results are consistent with the ones obtained for the
diluted ferromagnetic Ising model on a square lattice.

\end{abstract}
\pacs{PACS number(s): 05.20-y, 05.50+q, 64.60.Cn, 75.10.Hk, 75.40.Mg }
\keywords{Keywords: Ising model, Ferromagnetism, zero-temperature dynamics}

\begin{multicols}{2}

\narrowtext

\section{Introduction}
The Lenz-Ising model is probably the simplest
non-trivial model for cooperative behavior that shows spontaneous
breaking of symmetry . Because of this simplicity and the fact that
each individual element of the model modifies its behavior according
to the other individuals in its vicinity, it has a vast number of
applications ranging from solid-state physics to biology
\cite{Ising25,Temperley92,Binder}. An important extension to this
model is the introduction of disorder such as; random external fields,
random exchange parameters, and dilution, where only a fraction p of
the lattice sites are occupied by spins. In particular, the two
dimensional diluted ferromagnetic random Ising model is very
important to describe magnetic properties of several condensed matter
systems. Moreover, it constitutes a marginal situation of the Harris
criterion \cite{Harris74}, and for many years there has been several
theoretical and numerical studies
\cite{Dotsenko83,Shalaev94,Rodler98,Lima00} in order to clarify the
properties of the dilution model.

On the other hand, only in the last ten years the ``persistence''
problem in the randomic two-dimensional ferromagnetic Ising model has
attracted considerable interest\cite{Derrida94,Stauffer94,Majumdar98}.
In this most general form, this problem involves the fraction of space
which persists in its initial state until some time later. Hence, in
the non-equilibrium dynamics of spin systems we are interested in the
fraction of spins $P(t)$, that persist in the same state as at $t=0$
up to some later time $t$. For the pure ferromagnetic d-dimensional
$(d<4)$ Ising model, $P(t)$ has been found to decay
algebraically\cite{Derrida94,Bray94} \begin{equation} P(t) \propto
t^{-\theta (d)} \;\; , \end{equation}
where $\theta(d)$ is a non-trivial persistence exponent. For the same model
\cite{Stauffer94} at higher dimensions $(d>4)$ and for the two-dimensional
ferromagnetic $q$-state Potts model\cite{Derrida95} $(q>4)$ it has been shown
through computer simulations that $P(t)$ is not equal to zero for
$t \rightarrow \infty$. This characterictic is some time called as
``blocking''. Therefore, if $P(\infty) > 0$ we can reformulate the
problem by restricting our observation to those spins that eventually
flip. Hence, we can consider the behavior of
\begin{equation}
p(t) = P(t) - P(\infty)\;.
\end{equation}
By considering the dynamics of the local order parameter the persistence
problem can be generalized to non-zero temperatures\cite{Majumdar96,Zheng98}
Recently\cite{Jain99,Newman99} the attention has been turned to the
persistence problem in systems containing disorder. Numerical simulations
of the zero-temperature dynamics of the bond diluted (weak dilution
\cite{Jain99} or strong dilution\cite{Jain199}) two-dimensional Ising model
also reported ``blocking'' evidences. Howard\cite{Howard00} has found
evidence of an exponential decay of the persistence with blocking
for the homogeneous ferromagnetic Ising  model on the homogeneous tree of
degree three (T) with random spin configuration at time $0$.
Here we present results of an extensive numerical study of the
persistence of the ferromagnetic Ising model on the Voronoi-Delaunay
lattice. In this lattice, the coordination number and the distance
between the first neighbors sites is randomic. As the bond length
between the first neighbors varies randomly from neighbor to neighbor,
we consider that the coupling factor depends on the relative distance
$r_{ij}$ between sites $i$ and $j$ according to the following
expression: \begin{equation} J_{ij}=J_0 e^{-\alpha r_{ij}}
\end{equation}
where $J_0$ is a constant and $0 \le \alpha \le 1$.
The question to be answered here is: does the persistence behavior change
with this type of randomness or is its behavior the same as in the
diluted ferromagnetic Ising model in a regular lattice? In the present work,
we show that, with this type of randomness, $p(t)$ presents an
exponential decay as in the strongly diluted ferromagnetic Ising model
in two-dimension \cite{Newman99} in contrast with the behavior of the
pure and weakly diluted models.

\section{Model and simulation}

The Voronoi construction or tessellation for a given set of points in the
plane is defined as follows. For each point, we first determine the polygonal
cell consisting of the region of space nearer to that point than any other point.
Whenever two such cells share an edge, they are considered to be
neighbors. From the Voronoi tessellation, we can obtain the dual
lattice by the following procedure: when two cells are neighbors, a
link is placed between the two points located in the cells. From the
links, one obtains the triangulation of space that is called the
Delaunay lattice. The Delaunay lattice is dual to the Voronoy
tessellation in the sense that points correspond to cells, links to
edges and triangles to the vertices of the Voronoi tessellation.

We consider a two-dimensional Ising ferromagnetic on this Poissonian random
lattice which Hamiltonian is given by:
\begin{equation}
-K H=-\sum_{<i,j>} J_{ij} S_i S_j \; ,
\end{equation}
where $S_{i}= \pm 1$ are the Ising spins situated on every site of a
Delaunay lattice with $(LxL=N)$ sites and periodic boundary conditions;
$K=1/k_B T$ , $T$ is the temperature and $k_B$ is the Boltzmann constant.
The summation in Eq. $(4)$ runs over all nearest-neighbors pairs of sites
(points in the Delaunay construction).
In this lattice the coordination number varies locally between $3$ and
$\infty$ and the coupling factor $J_{ij}$ depends on the distance
between first neighbors according to Eq. $3$.

\begin{figure}

\includegraphics[width=8.0cm]{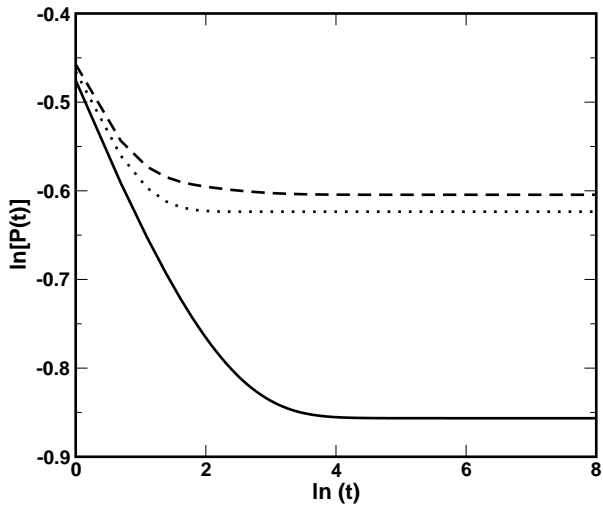}
\vspace{0.5cm}
\caption{$ln[P(t)]$ versus $ln(t)$ for three different values of the
parameter $\alpha$ and lattice size $L=500$.}
\end{figure}

For simplicity, the length of the system is defined here in terms of
the size of a regular lattice, $L=N^{1/2}$. We perform simulations
over a lattice with $L=500$. A randomly initial configuration of spins
is obtained and $P(t)$ is calculated over $6000$ of Monte Carlo steps
and a quenched average is done over 10 different Delaunay lattices for
each Monte Carlo step. The temperature zero Glauber dynamics was
utilized in order to check the number of spins that never change their
state at a time $t$. In this dynamics we start with a randomly
initial spin configuration and allow it to be updated by selecting one
spin to be flipped at random or following a given logical sequence.
The selected spin, $S_i$, is flipped or not according to $\Delta E_i$,
where $\Delta E_i$, is the energy of site $i$. If $\Delta E_i < 0$ the
spin $S_i$ is flipped with probability one. If $\Delta E_i = 0$ the
spin $S_i$ is flipped with probability $1/2$ chosen at random. Finally
if $\Delta E_i > 0$ the spin is not flipped. One Monte Carlo step
corresponds to application of the above rule for all spins of the
lattice. The system configuration is left to evolve until a given
Monte Carlo step $t \cong 6,000$. The number, $n(t)$, of sites that
do not change at this time $t$ is computed for each Monte Carlo step
for the determination of the persistence probability given
by\cite{Derrida94}:

\begin{figure}
\includegraphics[width=8.0cm]{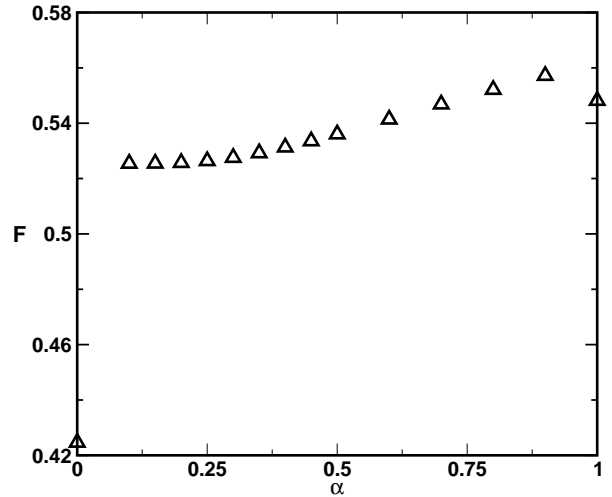}

\vspace{0.2cm}
\caption{Fraction of spins that never flip versus $\alpha$.}
\end{figure}

\begin{equation}
P(t)= \frac{<n(t)>}{N}
\end{equation}
The data have been obtained for 10 $\alpha$ values in the range
$0 \le \alpha \le 1$.

\section{Results and conclusion}
For $\alpha$ varying from $0$ (where the coupling constant is independent
of the distance between first neighbors) to $1$ the persistence $P(t)$
seems not to decay algebraically before the ``freezing'' as we can see
in Fig. $1$. In this figure we plot $ln\;P(t)$ versus $t$ for
$\alpha=0\;,\;0.5$ and $1$. We can also verify that, $P(t)=P(\infty)$
for $t>t^*$, where $t^*(\alpha)$ depends on the parameter $\alpha$,
growing with the $\alpha$ value. The fraction of frozen sites (i.e.
the sites that never flip) in function of the parameter $\alpha$ is
shown in Fig. $(2)$. This fraction has a maximum value near $\alpha =
0.9)$ decreasing for $(\alpha = 1)$. Finally in Fig. $(3)$ we plot
$ln\;[p(t)]$ versus $ln\;t$ for the same values of $\alpha$. From this
figure we can verify that $p(t)$ decay exponentially for long times.
This result agrees with the results obtained by Newman and
Stein\cite{Newman99} for the persistence in the strongly diluted Ising
ferromagnet. This behavior occurs for $\alpha=0$. This result is
contradictory for us once we know that for $\alpha=0$ the
ferromagnetic Ising model on a Delaunay lattice has the same critical
exponents that the pure ferromagnetic Ising model in a square lattice
has\cite{Lima00}. This fact does not agree with the persistence
behavior of the pure and weakly diluted ferromagnetic Ising model
reported by Jain \cite{Jain99}.

\begin{figure}
\includegraphics[width=8.0cm]{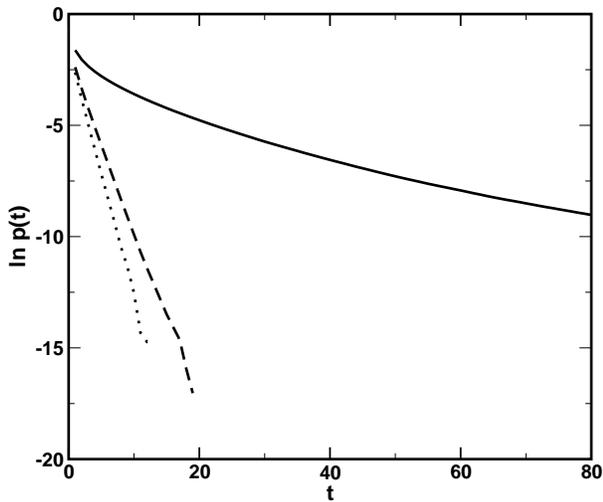}
\vspace{0.2cm}
\caption{A log-log plot of $p(t)$ versus $t$ for three
different $\alpha$.}
\end{figure}

In summary we have presented new data for the zero-temperature dynamics
of the two-dimensional ferromagnetic Ising model on a random Poissonian
lattice. This system exhibit ``blocking'' what means that the persistence
does not go to zero when $t \rightarrow \infty$. We also find that $p(t)$
decay exponentially in the long time regime. The fraction of ``frozen''
sites increases from a non-zero value with the parameter $\alpha$. Our
results strongly suggest that the persistence behavior is not algebraic
supporting the suggestion that the decay of the persistence probability
can be not-algebraic for certain classes of models.

\acknowledgments
This work was partially supported by CNPq (Brazilian research agencies).

\end{multicols}
\end{document}